\documentclass{www13-companion-accepted}

\usepackage[utf8]{inputenc}
\usepackage[T1]{fontenc}
\usepackage[T1,T2A]{fontenc}
\usepackage{lmodern}
\usepackage[activate=compatibility]{microtype}

\usepackage[hyphens]{url}
\usepackage[pdftex,urlcolor=black,colorlinks=true,linkcolor=black,citecolor=black]{hyperref}

\usepackage{enumitem}

\usepackage{color}

\usepackage{xspace}
\DeclareRobustCommand{\googleplus}{\mbox{Google\hspace{0em}\raisebox{.28ex}{\tiny\bf +}\kern-0.2ex}\xspace}

\usepackage{fancyvrb}
\usepackage{relsize}
\usepackage{listings}
\usepackage{verbatim}
\newcommand{\defaultlistingsize}{\fontsize{8pt}{9.5pt}}
\newcommand{\inlinelistingsize}{\fontsize{8pt}{11pt}}

\newcommand{\listingsize}{\defaultlistingsize}
\RecustomVerbatimCommand{\Verb}{Verb}{fontsize=\inlinelistingsize}
\RecustomVerbatimEnvironment{Verbatim}{Verbatim}{fontsize=\defaultlistingsize}
\let\oldurl\url
\renewcommand{\url}[1]{\inlinelistingsize\oldurl{#1}}

\lstdefinelanguage{JavaScript}{
  keywords={push, typeof, new, true, false, catch, function, return, null, catch, switch, var, if, in, while, do, else, case, break},
  keywordstyle=\bfseries,
  ndkeywords={class, export, boolean, throw, implements, import, this},
  ndkeywordstyle=\color{darkgray}\bfseries,
  identifierstyle=\color{black},
  sensitive=false,
  comment=[l]{//},
  morecomment=[s]{/*}{*/},
  commentstyle=\color{darkgray},
  stringstyle=\color{red},
  morestring=[b]',
  morestring=[b]"
}

\definecolor{grey}{RGB}{130,130,130}

\begin{document}

\conferenceinfo{World Wide Web Conference}{'13 Rio de Janeiro, Brazil}
\CopyrightYear{2013} 

\title{MJ no more: Using Concurrent Wikipedia Edit Spikes\\ with Social Network Plausibility Checks\\ for Breaking News Detection}

\numberofauthors{3}\author{
\alignauthor
Thomas Steiner\thanks{This work was partially supported by the European Commission
under Grant No.~248296 FP7 \mbox{I-SEARCH} project.}\\
	\affaddr{Google Germany GmbH}\\
	\affaddr{ABC-Str. 19}\\
	\affaddr{20354 Hamburg, Germany}\\
	\email{tomac@google.com} 
\alignauthor
Seth van Hooland\\
	\affaddr{Université Libre de Bruxelles}\\
	\affaddr{Avenue F.D. Roosevelt, 16}\\
	\affaddr{1050 Brussels, Belgium}\\
	\email{svhoolan@ulb.ac.be}	
\alignauthor
Ed Summers\\
	\affaddr{Library of Congress}\\
	\affaddr{101 Independence Ave, SE}\\
	\affaddr{Washington, DC 20540, USA}\\
	\email{ehs@pobox.com}	
}
\maketitle
\begin{abstract}
\fontencoding{T1}\selectfont
We have developed an application called \emph{Wikipedia Live Monitor}
that monitors article edits on different language versions of Wikipedia---%
as they happen in realtime.
Wikipedia articles in different languages are highly interlinked.
For example, the English article \emph{``en:2013\_Russian\_meteor\_event''}
on the topic of the February 15 meteoroid
that exploded over the region of Chelyabinsk Oblast, Russia,
is interlinked with \fontencoding{T2A}\selectfont
\emph{``ru:Падение\_метеорита\_на\_Урале\_в\_2013\_году''},
\fontencoding{T1}\selectfont
the Russian article on the same topic.
As we monitor multiple language versions of Wikipedia in parallel,
we can exploit this fact to detect \emph{concurrent edit spikes}
of Wikipedia articles covering the same topics,
both in only one, and in different languages.
We treat such concurrent edit spikes as signals
for potential breaking news events, whose plausibility we then check 
with full-text cross-language searches on multiple social networks.
Unlike the reverse approach of monitoring social networks first,
and potentially checking plausibility on Wikipedia second,
the approach proposed in this paper has the advantage of
being less prone to false-positive alerts, while being equally sensitive
to true-positive events, however, at only a~fraction of the processing cost.
A~live demo of our application is available online
at the URL \url{http://wikipedia-irc.herokuapp.com/},
the source code is available
under the terms of the Apache~2.0 license at 
\url{https://github.com/tomayac/wikipedia-irc}.

\end{abstract}




\pagebreak

\section{Introduction}

\subsection{Motivation}

Shortly after the celebrity news website TMZ
broke the premature news that the King of Pop \emph{Michael Jackson}~(MJ) had died,%
\footnote{MJ dead: \url{http://www.tmz.com/2009/06/25/michael-jackson-dies-death-dead-cardiac-arrest/},
accessed 02/18/2013}
the Internet slowed down.%
\footnote{Internet slow-down: \url{http://news.bbc.co.uk/2/hi/technology/8120324.stm},
accessed 02/18/2013}
Initially, Wikipedia's website administrators started noting abnormal load spikes~%
\cite{vibber2009currentevents}. Shortly afterwards, caching issues
caused by a~so-called edit war~\cite{beaumont2009editwar} led the site to go down:
Wikipedia editors worldwide made concurrent edits
to the Michael Jackson Wikipedia article, doing and undoing changes
regarding the tense of the article, death date,
and the circumstances of the (at the time) officially still unconfirmed fatality.
While Wikipedia engineers have worked hard
to ensure that future load spikes
do not take the site down again, there is without dispute a~lot of research potential
in analyzing such editing activity.

\subsection{Hypotheses and Research Questions}
\label{sec:hypotheses-and-research-questions}

In this paper, we present an application that monitors article edits
of different language versions of Wikipedia in realtime
in order to detect concurrent edit spikes that may be the source of
breaking news events.
When a~concurrent edit spike has been detected,
we use cross-language full-text searches on social networks
as plausibility checks to filter out false-positive alerts.
We are led by the following hypotheses.

\begin{itemize}
  \itemsep0em
  \item[(H1)] Breaking news events spread over social networks,
    independent from where the news broke initially.
  \item[(H2)] If a~breaking news event is important, it will be reflected on
    at least one language edition of Wikipedia.
  \item[(H3)] The time between when the news broke first and the news
    being reflected on Wikipedia is considerably short.   
\end{itemize}

\noindent These hypotheses lead us to the research questions below.

\begin{itemize}
  \itemsep0em
  \item[(Q1)] Can concurrent Wikipedia edit spikes combined with
    social network plausibility checks capture major breaking news events,
    and if so, with what delay?
  \item[(Q2)] Is the approach \emph{Wikipedia first, social networks second}
    at least as powerful as the reverse approach?
\end{itemize}

In this paper, we do not answer all research questions yet,
however, lay the foundation stone for future research in this area
by introducing the \emph{Wikipedia Live Monitor} application.

\section{Related Work}

We refer to an event as breaking news, if the event is of significant importance
to a~considerable amount of the population.
Petrovi\'{c} \emph{et~al.} define~\cite{petrovic2010streamingfirststory}
the goal of new event detection (or first story detection) as
\textit{``given a~sequence of stories, to identify the first story
to discuss a~particular event.''}
They define an event as \textit{``something that happens
at some specific time and place.''}
Classic streaming analysis of social network microposts so far has been mainly
focused on Twitter, a~microblogging social network that provides access
to a~sampled stream of generated microposts by means of its Streaming API.%
\footnote{Twitter Streaming API:
\url{https://dev.twitter.com/docs/api/1.1/get/statuses/sample},
accessed 02/18/2013}
Petrovi\'{c} \emph{et~al.} explain~\cite{petrovic2010streamingfirststory}:
\textit{``in the streaming model of computation,
items arrive continuously in a~chronological order, and have to be
processed in bounded space and time.''}
In the referenced paper, the authors report on a~system for streaming
new event detection applied to Twitter based on locality sensitive hashing.
Hu \emph{et~al.} provide an analysis of how news break and spread on Twitter~%
\cite{hu2012breakingnews}.
The task of linking news events with social media is covered by Tsagkias
\emph{et~al.} in~\cite{tsagkias2011linkingonlinenews}.
With this paper, we stand on the shoulders\footnote{Hence the title of this paper.}
of Osborne \emph{et~al.}~%
\cite{osborne2012bieber}, who use Wikipedia page view statistics%
\footnote{Page view statistics for Wikimedia projects:
\url{http://dumps.wikimedia.org/other/pagecounts-raw/},
accessed 02/18/2013}
as a~means to filter spurious events
stemming from event detection over social network streams.
Our approach reverses theirs, however, instead of the only hourly updated
page view statistics, we use realtime change notifications,
as explained in \autoref{sec:wikipedia-recent-changes}.
Our \emph{Wikipedia Live Monitor} is partly based on an application called
\emph{Wikistream}, developed by Ed Summers \emph{et~al.}, which was described in~%
\cite{summers2011odetonode}.

\begin{figure*}[t!]
  \centering
  \includegraphics[width=\linewidth]{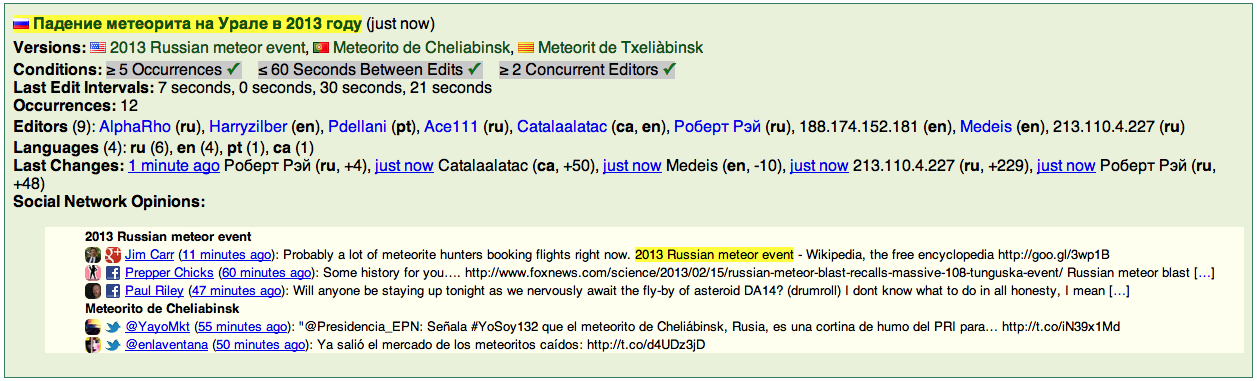}
  \caption{Screenshot with an article cluster of four concurrently edited articles
    (ru, en, pt, ca). All breaking news criteria are fulfilled,
    the cluster is a breaking news candidate.
    Cross-language social network search results for en and pt can be seen.}
  \label{fig:screenshot}
\end{figure*}

\section{Implementation Details}

\vspace{-0.5em}
\subsection{Wikipedia Recent Changes}
\label{sec:wikipedia-recent-changes}

As described earlier, our application monitors concurrent edit spikes
on different language versions of Wikipedia.
In the current implementation, we monitor 42 different Wiki\-pedias, 5 with 
$\geq$~1,000,000 and 37 with $\geq$~100,000 articles.%
\footnote{List of Wikipedias by size: \url{http://meta.wikimedia.org/wiki/List_of_Wikipedias},
accessed 02/18/2013}
Changes to any single one article are communicated by a~chat bot
over Wikipedia's own Internet Relay Chat~(IRC) server (\url{irc.wikimedia.org}),%
\footnote{Raw IRC feeds of recent changes: \url{http://meta.wikimedia.org/wiki/IRC/Channels\#Raw_feeds}, accessed 02/18/2013}
so that parties interested in the data can listen to the changes as they happen.
For each language version, there is a~specific chat room following the pattern
\texttt{"\#" + language + ".wikipedia"}.
For example, changes to Russian Wikipedia articles will be streamed to the room
\texttt{\#ru.wikipedia}.
A~special case is the room \texttt{\#wikidata.wikipedia} for Wikidata~%
~\cite{vrandecic2012wikidata},
a~platform for the collaborative acquisition and maintenance
of structured data to be used by
Wikimedia projects like Wikipedia.
A~sample chat message with the components separated
by the asterisk character \texttt{`*'}
announcing a~change can be seen in the following.
\texttt{"[[Juniata River]]
http://en.wikipedia.org/w/index.php?diff=516269072\&oldid=514\-659029 *
Johanna-Hypatia * (+67)
Category:Place names of Native American origin in Pennsylvania"}.
The message components are (i)~article name, (ii)~revision URL,
(iii)~Wikipedia editor handle, and (iv)~change size and change description.

\subsection{Article Clusters}

We cluster edits of articles about the same topic,
but written in different languages, in article clusters.
The example of the English
\emph{``en:2013\_Russian\_meteor\_event''}
and the Russian article \fontencoding{T2A}\selectfont
\emph{``ru:Падение\_метеорита\_на\_Урале\_в\_2013\_году''}
\fontencoding{T1}\selectfont that are both in the same cluster illustrate this.
We use the Wikipedia API to retrieve language links for a~given article.
The URL pattern for the API is as follows.
\url{http://$LANGUAGE.wikipedia.org/w/api.php?action=query&format=json&-prop=langlinks&titles=$ARTICLE}.

\subsection{Comparing Article Revisions}
\label{sec:comparing-article-revisions}

The Wikipedia API provides means to retrieve the actual changes
that were made during an edit including additions, deletions,
and modifications in a~\texttt{diff}-like manner.
The URL pattern is as follows.
\url{http://$LANGUAGE.wikipedia.org/w/api.php?action=compare&torev=$TO&fromrev=$FROM&format=json}. 
This allows us to classify edits in categories, like, \emph{e.g.},
negligible trivial edits (punctuation correction) and 
major important edits (new paragraph for an article),
which helps us to disregard seemingly concurrent edits
in order to avoid false-positive alerts.

\subsection{Breaking News Criteria}

Our application \emph{Wikipedia Live Monitor} puts  
detected article clusters in a~monitoring loop in which they remain
until their time-to-live (240~seconds) is over.
In order for an article cluster in the monitoring loop
to be identified as breaking news candidate,
the following breaking news criteria have to be fulfilled.

\begin{description}
  \itemsep0em
  \item[$\geq$~5~Occurrences:] An article cluster must have occurred
  in at least 5~edits.
  \item[$\leq$~60~Seconds Between Edits:] An article cluster may have
  at maximum 60~seconds in between edits.
  \item[$\geq$~2~Concurrent Editors:] An article cluster must have been edited
  by at least 2~concurrent editors.
  \item[$\leq$~240~Seconds Since Last Edit:] An article cluster's last edit
  may not be longer ago than 240~seconds.
\end{description}

The exact parameters of the breaking news criteria above
were \emph{determined empirically} by analyzing Wikipedia edits
over several hours and repeatedly adjusting the settings until
major news events happening at the same time were detected.
The resulting dataset split into three chunks has been made publicly available.%
\footnote{Wikipedia Live Monitor dataset: \url{https://www.dropbox.com/sh/2qsg1zhb8p35fxf/Dghn55y0kh},
accessed 02/18/2013}

\subsection{Social Network Plausibility Checks}

When a~breaking news candidate has been identified,
we use cross-language full-text social network searches 
on the social networks Twitter, Facebook, and \googleplus
as a~plausibility check.
As the \emph{article titles} of all language versions
of the particular article's cluster are know,
we use these very article titles as search queries for cross-language searches,
as can be seen in \autoref{fig:screenshot}.
This approach greatly improves the recall of the social network search,
however, requires either automatic translation, or an at least basic understanding
of the languages being searched in.
Currently the plausibility checking step is not yet fully automated,
as the search results are for the time being meant to be consumed by \emph{human evaluators}.
Driven by (H1), we assume breaking news events are being discussed on social networks.
We will show arguments for this assumption in \autoref{sec:premature-evaluation}.
For now, we expect social networks to be a~short period ahead of Wikipedia.
In consequence, if the human rater can find positive evidence
for a~connection between social network activities and Wikipedia edit actions,
the breaking news candidate is confirmed to indeed be breaking news.

\subsection{Application Pseudocode}

The \emph{Wikipedia Live Monitor} application has been implemented in Node.js,
a~server side JavaScript software system
designed for writing scalable Internet applications.
Programs are created using event-driven, asynchronous input/output operations
to minimize overhead and maximize scalability.
\autoref{code:pseudocode} shows the pseudocode of the two main event loops of the
\emph{Wikipedia Live Monitor} application.
The actual implementation is based on 
Martyn Smith's Node.js IRC library%
\footnote{Node.js IRC library: \url{https://github.com/martynsmith/node-irc},
accessed 02/18/2013} and
the WebSockets API and protocol~\cite{hickson2012websockets},
wrapped by  Guillermo Rauch's library Socket.IO.%
\footnote{Socket.IO library: \url{http://socket.io/},
accessed 02/18/2013}

\begin{lstlisting}[caption=The two main event loops
  of the application,
  label=code:pseudocode, float=b!]
§\textbf{Input: irc, listening on Wikipedia recent changes}§ 
§\textbf{Output: breakingNewsCandidates, breaking news candidates}§ 


monitoringLoop = {}
articleClusters = {}
breakingNewsCandidates = {}
 
§\textit{\# Event loop 1:}§
§\textit{\# When a new message arrives}§
irc.on.message §\textbf{do (article)}§
  langRefs = getLanguageReferences(article)
  articleRevs = getArticleRevisions(article)
  cluster = clusterArticles(article, langRefs)  
  
  §\textit{\# Create new cluster for previously unseen article}§
  §\textbf{if}§ cluster not in monitoringLoop
    monitoringLoop.push(cluster)
    articleClusters.push(cluster)
    updateStatistics(cluster)
    emit.newCluster(cluster, articleRevs)    
  §\textit{\# Update existing cluster, as the article was seen before}§
  §\textbf{else}§
    updateStatistics(cluster)
    emit.existingCluster(cluster, articleRevs)
    §\textit{\# Check breaking news criteria}§
    §\textbf{if}§ cluster.occurrences >= 5
      §\textbf{if}§ cluster.secsBetweenEdits <= 60      
        §\textbf{if}§ cluster.numEditors >= 2
          §\textbf{if}§ cluster.secsSinceLastEdit <= 240
            socialNetworks.search(langRefs)
            breakingNewsCandidates.push(cluster)
            emit.breakingNewsCandidate(cluster)            
          §\textbf{end if}§          
        §\textbf{end if}§  
      §\textbf{end if}§        
    §\textbf{end if}§          
  §\textbf{end if}§
  §\textbf{return}§ breakingNewsCandidates
§\textbf{end do}§

§\textit{\# Event loop 2:}§
§\textit{\# Remove too old clusters regularly}§
timeout.every.240seconds §\textbf{do}§
  §\textbf{for each}§ cluster §\textbf{in}§ monitoringLoop
    §\textbf{if}§ cluster.secsSinceLastEdit >= 240
      monitoringLoop.remove(cluster)
      articleClusters.remove(cluster)
    §\textbf{end if}§
  §\textbf{end for}§ 
§\textbf{end do}§
\end{lstlisting}

\section{Premature Evaluation}
\label{sec:premature-evaluation}

As noted earlier, in this paper,
we do not answer all research questions yet.
Nevertheless, we extract trends from our experiments so far.
In \autoref{sec:hypotheses-and-research-questions},
we have set up three hypotheses.
(H1) has been proven by Hu \emph{et~al.} in~\cite{hu2012breakingnews} for Twitter.
We argue that it can be generalized to other social networks
and invite the reader to have a~look at our dataset,
where the lively discussions about breaking news candidates
on the considered social networks Twitter, Facebook, and \googleplus
support the argumentation.
It is hard to prove (H2), as the concept of \emph{important breaking news}
is vague and dependent on one's personal background, however,
all evidence suggests that (H2) indeed holds true,
as, to the best of our knowledge and given our background,
what the authors consider \emph{important breaking news}
is represented on at least one language version of Wikipedia.
(H3) has been examined by Osborne \emph{et~al.} in~\cite{osborne2012bieber}.
In the paper, they suggest that Wikipedia lags about two hours behind Twitter.
It has to be noted that they look at hourly accumulated page (article) \emph{view} logs,
where we look at realtime article \emph{edit} log streams.
Our experiments suggest that the lag time of two hours
proposed by Osborne \emph{et~al.} may be too conservative. 
An educated guesstimation at this stage is that the lag time
for breaking news is more in the range of 30 minutes,
and for global breaking news like celebrity deaths
in the range of five minutes and less,
albeit the edits by our experience will be small and iterative
(\emph{e.g.}, ``X is a'' to ``X was a'', or the addition of a~death date),
followed by more consistent thorough edits.

The (at time of writing) recent breaking news event
of the resignation of \emph{Pope Benedict~XVI} helps respond to (Q1).
The three first edit times after the news broke on February 11, 2013
of the Pope's English Wikipedia article%
\footnote{Edit history en: \url{http://en.wikipedia.org/w/index.php?title=Pope_Benedict_XVI&action=history},
accessed 02/18/2013} 
are as follows
(all times in UTC): 10:58, 10:59, 11:02.
The edit times of the French article%
\footnote{Edit history fr: \url{http://fr.wikipedia.org/w/index.php?title=Beno\%C3\%AEt_XVI&action=history}, accessed 02/18/2013}
are as follows: 11:00, 11:00, 11:01.
This implies that by looking at only two language versions
(the actual number of monitored versions is 42) of the Pope article,
the system would have reported the news at 11:01.
The official Twitter account of Reuters announced%
\footnote{Reuters announces Pope resignation: \url{https://twitter.com/Reuters/status/300922108811284480},
accessed 02/18/2013} the news at 10:59.
Vatican Radio's announcement%
\footnote{Vatican Radio announces Pope resignation: \url{http://de.radiovaticana.va/Articolo.asp?c=663810},
accessed 02/18/2013} was made at 10:57:47.

Not all breaking news events have the same global impact as the Pope's resignation,
however, the proposed system was shown to work very reliably
also for smaller events of more regional impact, for example,
when \emph{Indian singer Varsha Bhosle} committed suicide%
\footnote{Varsha Bhosle suicide: \url{http://en.wikipedia.org/wiki/Varsha_Bhosle},
accessed 02/18/2013} on October 8, 2012.
A~systematic evaluation of (Q1)~compulsorily can only be done by random samples,
which has turned out positive results so far.
Again, we invite the reader to explore our dataset and to conduct own experiments.
A~systematic evaluation of (Q2) requires a~commonly shared dataset,
which we have provided, however, at this point in time, we do not have access to the system
of Osborne \emph{et~al.}

Regarding \emph{Wikipedia Live Monitor}'s scalability, we could scale the monitoring system
up to currently \emph{all} 284~Wikipedias on a~standard consumer laptop
(mid-2010 MacBook Pro, 2.66~GHz Intel Core~2, 8~GB RAM),
which once more proves the efficiency of the Node.js architecture
for this kind of event-driven applications.
In practice, however, the majority of the smaller Wikipedias being very rarely updated,
we limit ourselves to the Wikipedias with $\geq$~100,000 articles
at no remarkable loss of recall.

\section{Future Work}

Future work towards a~thorough scientific evaluation will mainly address two areas.
First, the \emph{automatic categorization of edits on Wikipedia}
needs to be more fine-grained.
In the context of breaking news detection, not all edits are equally useful.
An image being added to an article is an example of an edit
that usually will not be important.
In contrast, the category ``Living people'' being removed from an article
is a~strong indicator of breaking (sad) news.
Second, the \emph{connection between social network search and Wikipedia edits}
needs to be made clearer. 
In an initial step, the concrete changes to an article, as detailed in
\autoref{sec:comparing-article-revisions}, can be compared with
social network microposts using a~cosine similarity measure.
More advanced steps can exploit the potential knowledge from Wikipedia edits
(\emph{e.g.}, category ``Living people'' removed implies a~fatality).

\section{Conclusions}

In this paper, we have shown an application called \emph{Wikipedia Live Monitor}
and released its source code under the Apache~2.0 license.
This application monitors article edits on 42 different language versions of Wikipedia.
It detects breaking news candidates according to well-defined breaking news criteria,
whose exact parameters were determined empirically
and the corresponding dataset made available publicly.
We have shown how cross-language full-text social network searches are used
as plausibility checks to avoid false-positive alerts.
In a~first step towards a~full evaluation,
we have shown promising preliminary results
and actionable next steps in future work for improving the application.

\bibliographystyle{abbrv}
\bibliography{www2013devtrack}

\begin{thebibliography}{1}

\bibitem{beaumont2009editwar}
C.~Beaumont.
\newblock {Michael Jackson’s death sparks Wikipedia editing war}, June 2009.
\newblock
  \url{http://bit.ly/Michael-Jacksons-death-sparks-Wikipedia-editing-war},
  accessed 02/18/2013.

\bibitem{hickson2012websockets}
I.~Hickson.
\newblock {The WebSocket API}.
\newblock {Candidate Recommendation}, {W3C}, Sept. 2012.

\bibitem{hu2012breakingnews}
M.~Hu, S.~Liu, F.~Wei, Y.~Wu, J.~Stasko, and K.-L. Ma.
\newblock {Breaking News on Twitter}.
\newblock In {\em Proceedings of the 2012 ACM Annual Conference on Human
  Factors in Computing Systems}, CHI~'12, pages 2751--2754. ACM, 2012.

\bibitem{osborne2012bieber}
M.~Osborne, S.~Petrovi\'{c}, R.~McCreadie, C.~Macdonald, and I.~Ounis.
\newblock {Bieber no more: First Story Detection using Twitter and Wikipedia}.
\newblock In {\em Proceedings of the SIGIR Workshop on Time-aware Information
  Access}, 2012.

\bibitem{petrovic2010streamingfirststory}
S.~Petrovi\'{c}, M.~Osborne, and V.~Lavrenko.
\newblock {Streaming First Story Detection with Application to Twitter}.
\newblock In {\em Human Language Technologies: The 2010 Annual Conference of
  the North American Chapter of the Association for Computational Linguistics},
  HLT~'10, pages 181--189. Association for Computational Linguistics, 2010.

\bibitem{summers2011odetonode}
E.~Summers.
\newblock An ode to node, Nov. 2011.
\newblock \url{http://inkdroid.org/journal/2011/11/07/an-ode-to-node/},
  accessed 02/18/2013.

\bibitem{tsagkias2011linkingonlinenews}
M.~Tsagkias, M.~de~Rijke, and W.~Weerkamp.
\newblock {Linking Online News and Social Media}.
\newblock In {\em Proceedings of the Fourth ACM International Conference on Web
  Search and Data Mining}, WSDM~'11, pages 565--574. ACM, 2011.

\bibitem{vibber2009currentevents}
B.~Vibber.
\newblock Current events and traffic spikes, June 2009.
\newblock \url{http://blog.wikimedia.org/2009/06/25/current-events/}, accessed
  02/18/2013.

\bibitem{vrandecic2012wikidata}
D.~Vrande\v{c}i\'{c}.
\newblock {Wikidata: A New Platform for Collaborative Data Collection}.
\newblock In {\em Proceedings of the 21st International Conference Companion on
  World Wide Web}, WWW~'12 Companion, pages 1063--1064. ACM, 2012.

\end{thebibliography}

\end{document}